# Theoretical calculations on half-lives of spontaneous one-proton radioactivity*

WANG Hanlin[1], WANG Zhen[1], REN Zhongzhou[1, 2, *]

1. School of Physics Science and Engineering, Tongji University, Shanghai 200092, China
2. Key Laboratory of Advanced Micro-Structure Materials, Ministry of Education, Tongji University, Shanghai 200092, China

## Abstract

Research on the unstable nuclei beyond the nucleon drip line is an important method to study the nuclear interaction and structure in the extremely neutron-deficient or rich systems. Various nuclides beyond the proton drip line mainly decay through spontaneous one-proton emission. Using deformed Woods-Saxon potential, spin-orbit potential, and expanded Coulomb potential to construct the daughter-proton potential, the half-life data of various proton emitters are systematically calculated based on the quantum tunneling model and the microscopic Gamow state theory. By using nuclear data from different sources and comparing them with the measurements, the dependence of proton emission on decay energy and spectroscopic factors is evaluated. Additionally, based on previous observations, the half-life of the possibly lighter proton emitter in the fpg-shell below has been theoretically predicted. Our results are compiled into a comprehensive dataset of half-lives for both experimentally confirmed emitters ($50 < Z < 84$) and theoretically predicted emitters ($30 < Z < 50$), providing a useful reference for future experimental investigations related to the proton drip line. The datasets presented in this paper, including our results of calculation, are openly available at https://www.doi.org/10.57760/sciencedb.27551.

---

* The paper is an English translated version of the original Chinese paper published in Acta Physica Sinica. Please cite the paper as: WANG Hanlin[1], WANG Zhen[1], REN Zhongzhou[1, 2, *], **Theoretical calculations on half-lives of spontaneous one-proton radioactivity**. *Acta Phys. Sin.*, 2026, 75(2):020106. doi: 10.7498/aps.75.20251244

# 1 Introduction

As a complex quantum many-body system, nuclear stability depends on the numbers of protons and neutrons. When these numbers deviate significantly from the β-stability line, the nucleus can no longer bind the nucleons at the highest energy levels. These nucleons may then undergo spontaneous radioactive decay via nucleon emission. Consequently, the proton and neutron drip lines represent two natural boundaries of nuclear stability in nuclear structure studies. With advances in radioactive ion beam technology, experiments have progressively observed an increasing number of exotic nuclei near the drip lines. These exotic nuclei exhibit novel phenomena distinct from those near the valley of stability[1,2]. Compared with the neutron drip line, the proton drip line has been explored more extensively in experiments[3]. Therefore, proton radioactivity beyond the proton drip line has attracted considerable experimental and theoretical attention in recent years[4,5]. These rare proton-emitting nuclei possess negative one-proton separation energies, making proton radioactivity a spontaneous nuclear decay process. Investigating proton radioactivity data aids in understanding the physical properties of open quantum systems[6], shell evolution in drip-line regions[7], single-particle energy level structures and pairing correlations[8], and the rapid proton capture process[9].

Since the first experimental observation of one-proton radioactivity in $^{53}Co^{m}$ in 1970[10], an increasing number of one-proton emitters have been detected, including both ground and isomeric states, ranging from $^{53}Co^{m}$ to $^{185}Bi$ and spanning multiple shell regions[11]. On the theoretical side, the quantum tunneling model originally developed for α decay also applies to one-proton radioactivity. This model posits that the emitted proton resides in an extremely narrow resonance state within the parent nucleus-known as a Gamow state[12]. Due to quantum tunneling, the proton penetrates the potential barrier formed near the nuclear surface by the Coulomb and centrifugal potentials with a certain probability, escaping beyond the range of the nuclear interaction. To date, various interactions have been employed in studies of one-proton radioactivity, such as the cosh-type potential[13], single-folding M3Y interaction[14], and Skyrme interaction[15,16]. Concurrently, numerous theoretical models for calculating half-lives have been developed and applied to proton radioactivity, including the coupled-channel method[17] and relativistic density functional theory[18].

Recent studies have revealed that nuclear deformation plays a crucial role near the drip lines. For instance, theoretical investigations indicate that as nuclei approach the neutron drip line, their shapes gradually evolve toward ellipsoidal deformation until saturation occurs at the drip line, where further deformation becomes impossible[19]. Moreover, a recent phenomenological study demonstrated that incorporating deformation effects of both parent and daughter nuclei improves agreement between calculated results and experimental data in proton radioactivity[20]. Therefore, in the present theoretical model, we extend the original spherically symmetric potential by including quadrupole and hexadecapole deformations of the nucleus. We employ a more microscopic approach to calculate proton emission half-lives under deformation and quantitatively analyze the origins of certain nuclear data, particularly the influence of decay energy and spectroscopic factor choices on the computed results.

The remainder of this paper is organized as follows. Section 2 presents the theoretical model for deformed one-proton radioactivity, including the proton-daughter nucleus interaction potential and the half-life calculation method. Section 3 computes half-lives for experimentally observed proton emitters, provides theoretical predictions for lighter proton emitters within the *fpg* shell, and compiles a comprehensive dataset of half-lives that includes both currently known proton emitters ($30 < Z < 50$) and theoretically predicted ones ($50 < Z < 84$). Section 4 summarizes the conclusions of this work.

# 2 Theoretical Model for Deformed One-Proton Radioactivity

## 2.1 Proton-Daughter Nucleus Two-Body Interaction and Gamow States

Spontaneous one-proton radioactivity occurs in proton-unbound nuclei beyond the proton drip line, where the emitted proton tunnels through the Coulomb barrier via quantum tunneling and escapes from the parent nucleus. This mechanism resembles that of α decay[21-23], heavy cluster radioactivity[24,25], two-proton emission[26], and neutron radioactivity[27]. Thus, the emitted proton inside the nucleus can be treated as a Gamow state with an extremely narrow decay width[28]. By approximating the system as a two-body configuration consisting of the proton and the daughter nucleus, the proton in the Gamow state satisfies the time-independent Schrödinger equation.

$$\left[-\frac{\hbar^2}{2m_\mu}\frac{\mathrm{d}^2}{\mathrm{d}r^2} + V_{\mathrm{Nucl},\theta}(r) + V_{\mathrm{Coul},\theta}(r) + \frac{\hbar^2}{2m_\mu}\frac{\ell(\ell+1)}{r^2}\right] \times \chi_\theta(r) = E_0\chi_\theta(r), \tag{1}$$

Here, $\ell$ denotes the angular momentum transferred during the proton emission process, and $E_0$ represents the kinetic energy of the emitted proton. In the presence of deformation, the proton-daughter nucleus interaction potential depends on both the radial distance $r$ and the polar angle $\theta$ in spherical coordinates. For computational convenience, this study aligns the symmetry axis of the deformed daughter nucleus with the $z$-axis. The mass $m_\mu$ in the equation corresponds to the reduced mass of the two-body system.

The first term in the potential energy $V_{\mathrm{Nucl}}$, represents the nuclear interaction. To facilitate the incorporation of daughter nucleus deformation, this term is adopted in the Woods-Saxon form combined with a Thomas-form spin-orbit coupling potential[29]:

$$V_{\mathrm{Nucl},\theta}(r) = -\frac{V_{\mathrm{N0}}}{1+\exp\left[\frac{r-R_{\mathrm{N}}(\theta)}{a_{\mathrm{N}}}\right]} + \frac{\mathrm{d}}{r\mathrm{d}r}\frac{V_{\mathrm{so0}}\lambda_{\pi}^{2}}{1+\exp\left[\frac{r-R_{\mathrm{so}}(\theta)}{a_{\mathrm{so}}}\right]}\boldsymbol{\sigma}\cdot\boldsymbol{\ell}, \tag{2}$$

Here, $R_{\mathrm{N}}$ and $a_{\mathrm{N}}$ denote the radius and diffuseness parameters, respectively, of the daughter nucleus Woods-Saxon potential. $V_{\mathrm{so0}}$, $R_{\mathrm{so}}$, and $a_{\mathrm{so}}$ represent the strength, radius, and diffuseness parameters of the spin-orbit coupling potential, respectively. $\lambda_{\pi} \approx 1.414$ fm is the reduced Compton wavelength of the π pion[30]. Unlike spherical nuclei, in the deformed case, the radius parameters $R_{\mathrm{N}}$ and $R_{\mathrm{so}}$ are no longer isotropic constants. Instead, they are expressed via a multipole expansion in terms of spherical harmonics:

$$R_{\mathrm{N}}(\theta) = R_{\mathrm{N0}}[1+\beta_2 Y_{20}(\theta)+\beta_4 Y_{40}(\theta)] \tag{3}$$

where $\beta_2$ and $\beta_4$ are the quadrupole and hexadecapole deformation parameters of the daughter nucleus, respectively. The radius $R_{\mathrm{so}}$ can similarly be expanded based on the spherical radius $R_{\mathrm{so0}}$. The term $\boldsymbol{\sigma}\cdot\boldsymbol{\ell}$ represents the spin-orbit coupling of the emitted proton. Examination of the spin-orbit coupling term in the potential and the centrifugal term in the Schrödinger equation (1) reveals that the wave function of the emitted proton is closely associated with its orbital $n\ell_j$. Consequently, the characteristics of proton radioactivity exhibit a pronounced dependence on the proton orbital, making proton radioactivity a valuable tool for investigating the structure of nuclei beyond the proton drip line.

The spherical potential radii $R_{\mathrm{N0}}$ and $R_{\mathrm{so0}}$ can be determined using parameterized formulas. For the spin-orbit coupling potential, this study adopts the same parameters used in previous proton radioactivity investigations: $V_{\mathrm{so}} = 6.2$ MeV, $R_{\mathrm{so0}} = 1.01A_{\mathrm{D}}^{1/3}$ fm, and $a_{\mathrm{so}} = 0.75$ fm[30]. However, for the Woods-Saxon nuclear potential, since all proton-emitting nuclei examined in this work lie near the drip line and exhibit strong isospin asymmetry, we employ an isospin-dependent radius formula. This form has previously been validated for successfully calculating nuclear energy spectra and binding energies[31,32].

$$\begin{gathered} R_{\mathrm{N0}} = 1.19 A_{\mathrm{D}}^{1/3}(1 - 0.116\frac{N_{\mathrm{D}} - Z_{\mathrm{D}}}{A_{\mathrm{D}}}) + 0.235\mathrm{fm}, \\ a_{\mathrm{N}} = 0.637\mathrm{fm}. \end{gathered} \tag{4}$$

The second term in the potential energy is the Coulomb potential between the proton and the daughter nucleus. When the daughter nucleus is spherical, it can be approximated as a uniformly charged sphere with a charge distribution radius of $R_{\mathrm{C0}}$. In this case, the Coulomb potential admits an analytical expression derived from theoretical calculations. However, under deformation, the Coulomb potential cannot be obtained by simply adjusting the radius. Therefore, this study employs a multipole expansion method to compute the Coulomb potential for axially deformed nuclei[33].

$$V_{\mathrm{Coul},\theta}(r) = \frac{3Z_{\mathrm{D}}Z_{\mathrm{p}}e^2}{4\pi\varepsilon_0 R_{\mathrm{C0}}^3}\sum_{\lambda=0}^{\infty}\frac{4\pi Y_{\lambda 0}(\theta)}{2\lambda+1} \times \int_0^{\pi/2} Y_{\lambda 0}^*(\theta')K_\lambda(r,\theta')\sin\,\theta'\,\mathrm{d}\theta', \tag{5}$$

Here, $Z_{\mathrm{D}}$ and $Z_{\mathrm{p}}$ denote the atomic numbers of the daughter nucleus and the proton, respectively; $R_{\mathrm{C0}}$ represents the charge distribution radius in the spherical case; and $K_\lambda$ is a function of the deformed radius $R_{\mathrm{C}} = R_{\mathrm{C0}}[1 + \beta_2 Y_{20}\theta + \beta_4 Y_{40}(\theta)]$:

$$K_\lambda = \begin{cases} \dfrac{(2\lambda+1)r^2}{(\lambda+3)(\lambda-2)} - \dfrac{r^\lambda(\lambda-2)^{-1}}{R_{\mathrm{C}}^{\lambda-2}(\theta)}, r \leqslant R_{\mathrm{C}}(\theta), \lambda \neq 2, \\ \dfrac{r^2}{5} + r^2\ln\left(\dfrac{R_{\mathrm{C}}(\theta)}{r}\right), r \leqslant R_{\mathrm{C}}(\theta), \lambda = 2, \\ \dfrac{1}{\lambda+3}\dfrac{R_{\mathrm{C}}^{\lambda+3}(\theta)}{r^{\lambda+1}}, r > R_{\mathrm{C}}(\theta). \end{cases} \tag{6}$$

The deformed Coulomb potential can be calculated using the two equations above, with $\lambda$ taken as an even integer. In this work, the Coulomb radius adopts a form with fewer parameters: $R_{\mathrm{C0}} = 1.2247 A_{\mathrm{D}}^{1/3}$[34].

The preceding analysis specifies all potential parameters except $V_{\mathrm{N0}}$. According to theoretical considerations, the depth $V_{\mathrm{N0}}$ should be chosen such that the real part of the energy, $n\ell_j$, of the corresponding Gamow-state proton orbital $E_0$ -which corresponds to the kinetic energy of the emitted proton-satisfies the relation $E_0 = QA_{\mathrm{D}}/(A_{\mathrm{D}}+1)$ with the experimentally measured decay energy $Q$, where $A_{\mathrm{D}}$ denotes the nucleon number of the daughter nucleus. The Gamow states are obtained by solving the time-independent Schrödinger equation, namely Equation (1). However, unlike bound-state solutions, the Gamow-state wavefunctions must satisfy boundary conditions that asymptotically match outgoing Coulomb wavefunctions at large $r$, since the Coulomb potential dominates the total interaction in this region[17].

$$\lim_{r\to\infty}\chi(r) = N[G_\ell(kr) + \mathrm{i}F_\ell(kr)], \tag{7}$$

Here, $G_\ell$ and $F_\ell$ denote the irregular and regular Coulomb wave functions, respectively, and $N$ represents the normalization factor. Imposing this boundary condition yields the energy and wave function of the Gamow state for a given potential depth $V_{\mathrm{N0}}$. After adjusting $V_{\mathrm{N0}}$ to reproduce the experimentally observed $E_0$, all parameters in the two-body potential are fully determined. Detailed numerical results will be thoroughly analyzed in the computational section.

## 2.2 Half-life calculation under deformation

Once the proton-daughter nucleus interaction potential is established, the half-life for proton emission via quantum tunneling from the Gamow state to the exterior of the nucleus can be computed using several approaches. The semiclassical WKB approximation estimates the tunneling probability through the Coulomb barrier but does not incorporate the wave function information obtained in Section 2.1. Therefore, this study employs the amplitude of the Gamow-state wave function in the exterior region to calculate the half-life. Compared with semiclassical methods, this approach offers a more microscopic description. Based on the computed wave function, the decay width in the $\theta$ direction can be expressed as[35]

$$\varGamma(\theta) = \frac{\hbar^2 k}{m_\mu}\frac{|\chi_\theta(r_\mathrm{b})|^2}{G_\ell^2(kr_\mathrm{b}) + F_\ell^2(kr_\mathrm{b})}, \tag{8}$$

Here, $k = \sqrt{2m_\mu E_0}/\hbar$ denotes the wave number of the emitted proton, and $r_\mathrm{b}$ represents a sufficiently large distance beyond the range of the nuclear potential. Combining this with the boundary condition in Equation (7), we observe that when $r_\mathrm{b}$ is sufficiently large, the results become insensitive to the specific choice of $r_\mathrm{b}$. The total decay width is obtained by averaging the partial widths over all azimuthal angles in spherical coordinates:

$$\overline{\varGamma} = \frac{1}{2}\int_0^{\pi}\varGamma(\theta)\sin\,\theta\,\mathrm{d}\theta. \tag{9}$$

The half-life can subsequently be calculated using the relation $T_{1/2} = \hbar\ln\,2/(S_\mathrm{p}\overline{\varGamma})$, where $S_\mathrm{p}$ denotes the spectroscopic factor, which can be computed via methods such as BCS theory. In this theoretical framework, when nuclear deformation is neglected, the spectroscopic factor physically represents the probability that the proton orbital $n\ell_j$ in the daughter nucleus remains unoccupied[36]. When deformation of either the parent or daughter nucleus is considered, the magnitude of the spectroscopic factor remains related to this probability but requires deformation-dependent corrections[37].

# 3 Results and Discussion

Based on the theoretical model outlined in Section 2, this section presents systematic calculations for several proton-emitting nuclei recently identified in experiments. The computed results are then compared with experimental half-life values[3,38]. Nuclear data used in this work are taken from the NUBASE 2020 evaluation[38]. Quadrupole and hexadecapole deformation parameters are adopted from FRDM calculations[39]. The decay energy $Q$, employed to compute the proton emission resonance energy $E_0$, primarily relies on experimental one-proton separation energies derived from the AME 2020 atomic mass evaluation[40,41] as listed in the NNDC database[3], supplemented by FRDM theoretical values[39]. For nuclei lacking precise experimental $Q$-values, we instead use decay energies measured directly in relevant proton radioactivity experiments; the corresponding references are indicated in the data tables of this paper. Spectroscopic factors are mainly obtained from RMF (Relativistic Mean Field) theoretical calculations[42] and deformation-dependent fitted values[13]. Exceptions are explicitly noted in Table 1.

**Table 1 Experimental data of various proton emitters, where the orbital is occupied in the spherical limit. $Q_{\text{exp}}$ and $Q_{\text{FRDM}}$ represent the decay energies derived from experiments (obtained from NNDC and AME 2020[3,40,41], except for values explicitly marked otherwise) and the Finite-Range Droplet Model (FRDM) theory[39], respectively. $\beta_{2,4}$ denote the deformation parameters, $T_{1/2}^{\text{exp}}$ is the experimental half-life, and the last two columns present the spectroscopic factors calculated using the Relativistic Mean Field (RMF) model and deformation-dependent fitting.**

| Parent Nuclei | $n\ell_j$ | $Q_{\text{exp}}$/MeV | $Q_{\text{FRDM}}$/MeV | $\beta_2$ | $\beta_4$ | $\log_{10} T_{1/2}^{\text{exp}}$ | $S_{\text{p}}^{\text{RMF}}$ | $S_{\text{p}}^{\text{fit}}$ |
|---|---|---|---|---|---|---|---|---|
| $^{109}$I | $2d_{5/2}$ | 0.820 | 0.821 | 0.139 | 0.056 | −4.029 | 0.726 | 0.099 |
| $^{112}$Cs | $2d_{5/2}$ | 0.816 | 0.901 | 0.185 | 0.052 | −3.310 | 0.369 | 0.063 |
| $^{113}$Cs | $2d_{3/2}$ | 0.973 | 0.681 | 0.195 | 0.054 | −4.752 | 0.373 | 0.057 |
| $^{117}$La | $2d_{3/2}$ | 0.820 | 0.581 | 0.282 | 0.106 | −1.602 | 0.311 | 0.024 |
| $^{121}$Pr | $2d_{3/2}$ | 0.890 | 0.671 | 0.304 | 0.087 | −2.000 | 0.122 | 0.019 |
| $^{130}$Eu | $2d_{3/2}$ | 1.028[45] | 1.111 | 0.331 | 0.018 | −3.046 | 0.816 | 0.014 |
| $^{131}$Eu | $2d_{3/2}$ | 0.947 | 0.961 | 0.331 | 0.018 | −1.699 | 0.029 | 0.014 |
| $^{135}$Tb | $2f_{7/2}$ | 1.188 | 1.131 | 0.322 | −0.037 | −3.027 | 0.028 | 0.016 |
| $^{140}$Ho | $2f_{7/2}$ | 1.094 | 0.881 | 0.276 | −0.047 | −2.222 | 0.952 | 0.025 |
| $^{141}$Ho | $2f_{7/2}$ | 1.177 | 0.761 | 0.253 | −0.039 | −2.387 | 0.008 | 0.032 |
| $^{141}$Ho$^{m}$ | $3s_{1/2}$ | 1.243 | 0.827 | 0.253 | −0.039 | −5.137 | 0.048 | 0.032 |
| $^{144}$Tm | $1h_{11/2}$ | 1.712 | 1.201 | 0.254 | −0.064 | −5.721 | 0.558[43] | 0.031 |
| $^{145}$Tm | $1h_{11/2}$ | 1.736 | 1.141 | 0.231 | −0.068 | −5.499 | 0.580 | 0.039 |
| $^{146}$Tm | $1h_{11/2}$ | 1.196 | 0.951 | 0.219 | −0.057 | −1.125 | 0.962 | 0.045 |
| $^{146}$Tm$^{m}$ | $1h_{11/2}$ | 1.120 | 0.833 | 0.219 | −0.057 | −0.703 | 0.962 | 0.045 |
| $^{147}$Tm | $1h_{11/2}$ | 1.059 | 0.661 | −0.187 | −0.022 | 0.587 | 0.581 | 0.187 |
| $^{147}$Tm$^{m}$ | $2d_{3/2}$ | 1.127 | 0.729 | −0.187 | −0.022 | −3.444 | 0.953 | 0.187 |
| $^{150}$Lu | $1h_{11/2}$ | 1.270 | 1.001 | −0.167 | −0.035 | −1.197 | 0.497 | 0.199 |
| $^{151}$Lu | $1h_{11/2}$ | 1.241 | 1.001 | −0.167 | −0.035 | −0.896 | 0.490 | 0.199 |
| $^{151}$Lu$^{m}$ | $2d_{3/2}$ | 1.319 | 1.079 | −0.167 | −0.035 | −4.796 | 0.858 | 0.199 |
| $^{155}$Ta | $1h_{11/2}$ | 1.453 | 1.191 | 0.021 | 0.000 | −2.538 | 0.422 | 0.324 |
| $^{156}$Ta | $2d_{3/2}$ | 1.020 | 0.591 | −0.063 | 0.001 | −0.842 | 0.761 | 0.274 |
| $^{156}$Ta$^{m}$ | $1h_{11/2}$ | 1.122 | 0.693 | −0.063 | 0.001 | 0.933 | 0.493 | 0.274 |
| $^{157}$Ta | $3s_{1/2}$ | 0.935 | 0.771 | −0.084 | 0.014 | −0.527 | 0.797 | 0.257 |
| $^{159}$Re$^{m}$ | $1h_{11/2}$ | 1.720 | 1.301 | 0.085 | 0.003 | −4.665 | 0.387[43] | 0.171 |
| $^{160}$Re | $2d_{3/2}$ | 1.267 | 0.821 | 0.107 | 0.004 | −3.045 | 0.507 | 0.137 |
| $^{161}$Re | $3s_{1/2}$ | 1.197 | 0.761 | 0.128 | 0.018 | −3.357 | 0.892 | 0.111 |
| $^{161}$Re$^{m}$ | $1h_{11/2}$ | 1.321 | 0.885 | 0.128 | 0.018 | −0.678 | 0.290 | 0.111 |
| $^{164}$Ir$^{m}$ | $1h_{11/2}$ | 1.814 | 1.928 | 0.118 | −0.007 | −4.155 | 0.339[43] | 0.123 |
| $^{165}$Ir$^{m}$ | $1h_{11/2}$ | 1.740 | 1.311 | 0.129 | 0.006 | −3.462 | 0.187 | 0.110 |
| $^{166}$Ir | $2d_{3/2}$ | 1.152 | 0.711 | 0.140 | −0.005 | −0.824 | 0.415 | 0.098 |
| $^{166}$Ir$^{m}$ | $1h_{11/2}$ | 1.324 | 0.883 | 0.140 | −0.005 | −0.076 | 0.188 | 0.098 |
| $^{167}$Ir | $3s_{1/2}$ | 1.070 | 0.731 | 0.151 | −0.004 | −1.028 | 0.912 | 0.088 |
| $^{167}$Ir$^m$ | $1h_{11/2}$ | 1.245 | 0.906 | 0.151 | −0.004 | 0.848 | 0.183 | 0.088 |
| $^{170}$Au | $2d_{3/2}$ | 1.472 | 0.981 | 0.129 | 0.007 | −3.493 | 0.511 | 0.110 |
| $^{170}$Au$^m$ | $1h_{11/2}$ | 1.757 | 1.266 | 0.129 | 0.007 | −2.973 | 0.137[43] | 0.110 |
| $^{171}$Au | $3s_{1/2}$ | 1.448 | 0.961 | 0.129 | −0.006 | −4.770 | 0.848 | 0.110 |
| $^{171}$Au$^m$ | $1h_{11/2}$ | 1.707 | 1.220 | 0.129 | −0.006 | −2.654 | 0.087 | 0.110 |
| $^{176}$Tl | $3s_{1/2}$ | 1.265 | 1.031 | −0.115 | −0.03 | −2.284 | 0.926 | 0.233 |
| $^{177}$Tl | $3s_{1/2}$ | 1.156 | 0.981 | −0.115 | −0.03 | −1.176 | 0.733 | 0.233 |
| $^{177}$Tl$^m$ | $1h_{11/2}$ | 1.963 | 1.788 | −0.115 | −0.03 | −3.346 | 0.022 | 0.233 |
| $^{185}$Bi | $3s_{1/2}$ | 1.598[46] | 1.611 | 0.000 | 0.012 | −4.191 | 0.011 | 0.400 |

## 3.1 Calculated proton radioactivity half-life data for nuclei in the region 50 < $Z$ < 84

First, prior to computing data for a large number of proton-emitting nuclei, we selected $^{151}$Lu-one of the first ground-state proton emitters observed experimentally-as a representative case to illustrate the computational procedure and detailed results, thereby validating the consistency between our theoretical model and experimental findings. The ground state of $^{151}$Lu consists of an even-even daughter nucleus $^{150}$Yb and a proton occupying the Gamow state in the $1h_{11/2}$ orbital under the spherical limit. Among all its decay branches, proton radioactivity constitutes the dominant component, with a branching ratio of 63.4%[3]. The experimental decay energy is 1.241 MeV[3,40,41], yielding an experimentally determined proton radioactivity half-life of 127 ms. For clarity, we computed the proton resonance state in this orbital using the theoretical model summarized in Section 2, neglecting nuclear deformation, as shown in Figure 1. In the figure, the black, blue, and red solid lines, along with the yellow dashed line, represent the total potential, nuclear potential, Coulomb potential, and the real part of the proton Gamow state energy $E_0$, respectively, corresponding to the left vertical axis. The green and purple dashed lines depict the real and imaginary parts of the calculated proton Gamow state wave function. Although the wave function exhibits oscillatory behavior in the external region beyond the nucleus due to the boundary conditions in Equation (7), it remains predominantly localized within the parent nucleus, inside the Coulomb barrier.

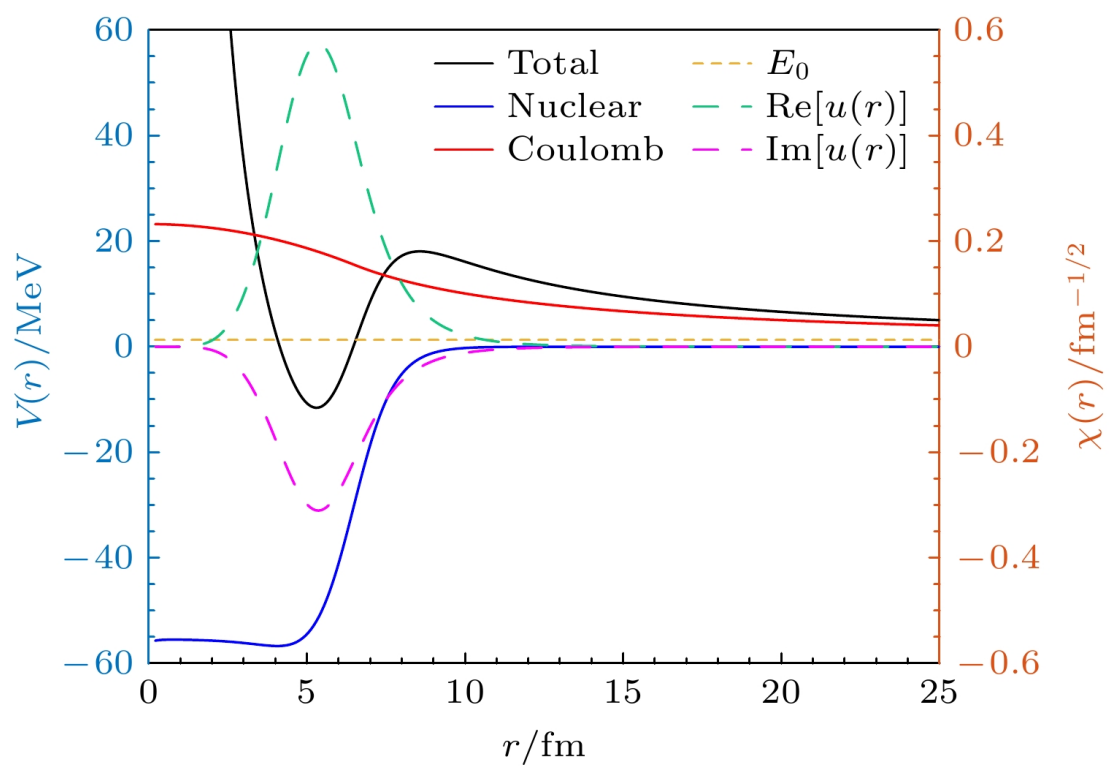

**Fig. 1 The daughter-proton interaction potential in $^{151}$Lu for the spherical case and the Gamow state wave function of the emitted proton. The solid lines denote different components of the potential, while the real part (green dashed line) and the imaginary part (purple dashed line) of the wave function correspond to the right axis, with the real part of the energy indicated by the yellow dotted horizontal line.**

Subsequently, the quadrupole and hexadecapole deformations of the daughter nucleus $^{150}$Yb were further considered. Using the deformation parameters $\beta_2 = -0.167$ and $\beta_4 = -0.035$ from the Finite-Range Droplet Model (FRDM)[39] calculations, the decay widths along various directions were computed by adjusting the nuclear radius in each direction. The half-life was then obtained by averaging these directional decay widths. Figure 2 illustrates the deformation of the daughter nucleus and the corresponding angular dependence of the decay width $\Gamma(\theta)$. The blue dashed line represents the angular-dependent radius $R_{\rm N}(\theta)$ of the Woods-Saxon potential, while the purple solid line depicts the angular distribution of the decay width. Since $^{150}$Yb is an oblate nucleus with $\beta_2 < 0$, its radius is smaller along the symmetry axis, resulting in a narrower decay width. Conversely, both the radius and the decay width reach their maximum values in the direction perpendicular to the symmetry axis. After performing the angular averaging and incorporating the spectroscopic factor $S_{\rm p} =$ 0.49 derived from Relativistic Mean-Field (RMF) theory, the calculated half-life under deformation is 70.2 ms, which shows reasonable agreement with the experimental value.

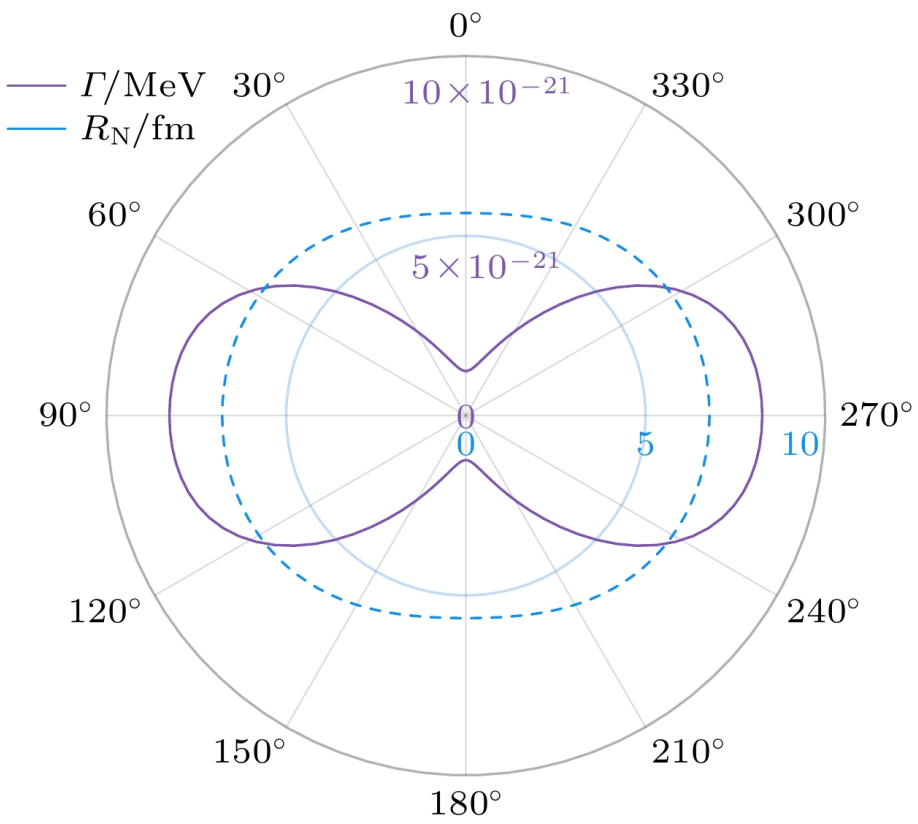

**Fig. 2 The angular distribution of the deformed radius (dashed line) and the decay width (solid line).**

After validating the theoretical model, we applied it to calculate the half-lives of numerous proton-emitting nuclei in the region $50 < Z < 84$, where most experimentally observed proton emitters are concentrated. Table 1 lists relevant data for these nuclei, including the proton orbital, deformation parameters of the daughter nucleus, and experimental half-life values. To analyze the dependence of proton emission on decay energy and spectroscopic factor data, different datasets were employed in the calculations presented in Table 1. Specifically, for decay energy, we used both the experimental values $Q_{\mathrm{exp}}$[3] and the one-proton separation energies $Q_{\mathrm{FRDM}}$[39] derived from the Finite-Range Droplet Model (FRDM) nuclear ground-state mass table. For the spectroscopic factor, we utilized two sets of data: results from Relativistic Mean-Field (RMF) theory, denoted as $S_{\mathrm{p}}^{\mathrm{RMF}}$[42,43], and values obtained from a deformation-dependent fitting formula, denoted as $S_{\mathrm{p}}^{\mathrm{fit}}$[13]. Columns 2 through 5 in Table 2 present several sets of theoretical half-life calculations from this work. Given the substantial variation in half-life magnitudes across different proton emitters, logarithmic values of the half-lives are reported in the table. To facilitate a more intuitive comparison between theoretical results and experimental data, we computed the difference between each calculated half-life and its corresponding experimental value under various parameter sets; these differences are displayed in Figure 3. In this figure, points with different colors and shapes represent half-life calculations using distinct combinations of decay energy and spectroscopic factor data. Specifically, Figure 3(a) corresponds to spectroscopic factors derived from the fitting formula, whereas Figure 3(b) uses spectroscopic factors from RMF theory. The yellow-shaded region indicates cases where the discrepancy between calculated and experimental values falls within one order of magnitude. Under both spectroscopic factor scenarios, when employing the one-proton separation energies calculated from FRDM, the root-mean-square deviation between the computed half-lives and experimental values

$$\sigma = \sqrt{\frac{\sum(\log_{10}\ T_{1/2} - \log_{10}\ T_{1/2}^{\exp})^2}{N}}$$

The root-mean-square (RMS) deviations are $\sigma(Q_{\mathrm{FRDM}}, S_{\mathrm{p}}^{\mathrm{RMF}}) = 4.90$ and $\sigma(Q_{\mathrm{FRDM}}, S_{\mathrm{p}}^{\mathrm{fit}}) = 5.38$, respectively. When experimental decay energies are employed, the RMS deviations for the two spectroscopic factors become $\sigma(Q_{\exp}, S_{\mathrm{p}}^{\mathrm{RMF}}) = 0.46$ and $\sigma(Q_{\exp}, S_{\mathrm{p}}^{\mathrm{fit}}) = 0.66$, respectively. Under these conditions, the theoretical model adopted in this work enables relatively accurate calculations of proton radioactivity half-lives. Furthermore, the last two columns of Table 2 present comparisons of these results with those obtained from two other theoretical models: an analytical semi-classical approach[11] and an extended Geiger-Nuttall law formula[44]. Figure 4 displays the differences between the logarithms of the experimental half-lives and the results from these two models, as well as from our model using experimental decay energies and RMF spectroscopic factors. The RMS deviations of these two sets of results relative to experimental values are 0.60 and 0.38, respectively-comparable to our RMS deviation of $\sigma(Q_{\exp}, S_{\mathrm{p}}^{\mathrm{RMF}}) = 0.46$. Consequently, all these theoretical models can achieve reasonably precise predictions of proton radioactivity half-lives.

**Table 2 Data calculated by our theoretical model, using different spectroscopic factors and decay energies listed in Table 1. The second and third columns correspond to the experimental decay energies[3,40,41], whereas the fourth and fifth columns correspond to the decay energies derived from the Finite-Range Droplet Model (FRDM)[39], denoted as $Q_{\mathrm{FRDM}}$. The last two columns present results from other theoretical studies.**

| Parent Nuclei | Based on $Q_{\exp}$ | | Based on $Q_{\mathrm{FRDM}}$ | | Other works | |
|---|---|---|---|---|---|---|
| | $\log_{10}\ T_{1/2}^{\mathrm{RMF}}$ | $\log_{10}\ T_{1/2}^{\mathrm{fit}}$ | $\log_{10}\ T_{1/2}^{\mathrm{RMF}}$ | $\log_{10}\ T_{1/2}^{\mathrm{fit}}$ | Ref. [11] | Ref. [44] |
| $^{109}$I | −4.673 | −3.809 | −4.688 | −3.824 | −4.098 | −3.507 |
| $^{112}$Cs | −3.587 | −2.816 | −4.823 | −4.053 | −3.261 | −2.844 |
| $^{113}$Cs | −5.746 | −4.928 | −1.171 | −0.353 | −5.599 | −4.796 |
| $^{117}$La | −2.884 | −1.765 | 2.101 | 3.219 | −2.699 | −2.072 |
| $^{121}$Pr | −2.840 | −2.032 | 1.147 | 1.955 | −3.020 | −2.552 |
| $^{130}$Eu | −4.218 | −2.468 | −5.211 | −3.461 | −3.470 | −3.121 |
| $^{131}$Eu | −1.680 | −1.379 | −1.876 | −1.575 | −2.354 | −2.141 |
| $^{135}$Tb | −3.279 | −3.033 | −2.652 | −2.405 | −3.647 | −3.380 |
| $^{140}$Ho | –3.123 | –1.545 | –0.031 | 1.547 | –1.909 | –1.902 |
| $^{141}$Ho | –2.013 | –2.611 | 4.341 | 3.743 | –2.878 | –2.811 |
| $^{141}$Ho$^{m}$ | –5.188 | –5.007 | 0.549 | 0.729 | –5.588 | –5.783 |
| $^{144}$Tm | –5.435 | –4.185 | –1.044 | 0.206 | –4.686 | –5.216 |
| $^{145}$Tm | –5.603 | –4.436 | –0.355 | 0.812 | –4.835 | –5.401 |
| $^{146}$Tm | –1.214 | 0.120 | 2.058 | 3.393 | –0.204 | –1.272 |
| $^{146}$Tm$^{m}$ | –0.313 | 1.022 | 4.129 | 5.463 | 0.702 | –0.999 |
| $^{147}$Tm | 0.725 | 1.218 | 8.322 | 8.814 | 0.408 | 0.681 |
| $^{147}$Tm$^{m}$ | –3.591 | –2.884 | 3.079 | 3.786 | –3.728 | –2.455 |
| $^{150}$Lu | –1.148 | –0.750 | 2.250 | 2.647 | –1.530 | –1.199 |
| $^{151}$Lu | –0.844 | –0.452 | 2.248 | 2.639 | –1.211 | –0.911 |
| $^{151}$Lu$^{m}$ | –5.053 | –4.418 | –2.304 | –1.669 | –5.213 | –3.899 |
| $^{155}$Ta | –2.318 | –2.204 | 1.191 | 0.509 | –2.764 | –2.397 |
| $^{156}$Ta | –0.986 | –0.542 | 8.836 | 9.280 | –0.901 | –0.180 |
| $^{156}$Ta$^{m}$ | 1.177 | 1.433 | 9.178 | 9.434 | 0.839 | 1.101 |
| $^{157}$Ta | –0.299 | 0.193 | 2.923 | 3.416 | –0.505 | –0.797 |
| $^{159}$Re$^{m}$ | –3.966 | –3.610 | –0.290 | 0.066 | –4.399 | –4.586 |
| $^{160}$Re | –3.136 | –2.568 | 3.681 | 4.250 | –3.438 | –2.450 |
| $^{161}$Re | –3.439 | –2.533 | 3.901 | 4.807 | –3.575 | –3.277 |
| $^{161}$Re$^{m}$ | –0.415 | 0.003 | 5.771 | 6.189 | –0.910 | –0.729 |
| $^{164}$Ir$^{m}$ | –4.152 | –3.711 | –4.890 | –4.448 | –4.577 | –4.247 |
| $^{165}$Ir$^{m}$ | –3.392 | –3.161 | 0.413 | 0.645 | –4.051 | –3.550 |
| $^{166}$Ir | –1.080 | –0.454 | 7.189 | 7.814 | –1.386 | –0.801 |
| $^{166}$Ir$^{m}$ | 0.255 | 0.537 | 6.672 | 6.953 | –0.370 | –0.344 |
| $^{167}$Ir | –1.142 | –0.126 | 5.450 | 6.465 | –1.186 | –1.347 |
| $^{167}$Ir$^{m}$ | 1.142 | 1.460 | 6.213 | 6.530 | 0.532 | 0.546 |
| $^{170}$Au | –4.130 | –3.462 | 2.053 | 2.720 | –4.331 | –3.254 |
| $^{170}$Au$^{m}$ | –2.954 | –2.857 | 1.592 | 1.688 | –3.693 | –3.330 |
| $^{171}$Au | –4.954 | –4.066 | 1.332 | 2.220 | –5.028 | –4.460 |

| $^{171}Au^m$ | −2.390 | −2.491 | 2.344 | 2.243 | −3.313 | −2.992 |
|---|---|---|---|---|---|---|
| $^{176}Tl$ | −2.470 | −1.871 | 0.799 | 1.397 | −2.451 | −2.361 |
| $^{177}Tl$ | −0.978 | −0.481 | 1.742 | 2.239 | −1.035 | −1.263 |
| $^{177}Tl^m$ | −3.149 | −4.174 | −1.962 | −2.988 | −4.611 | −3.543 |
| $^{185}Bi$ | −3.383 | −4.944 | −3.495 | −5.055 | −5.214 | −4.730 |

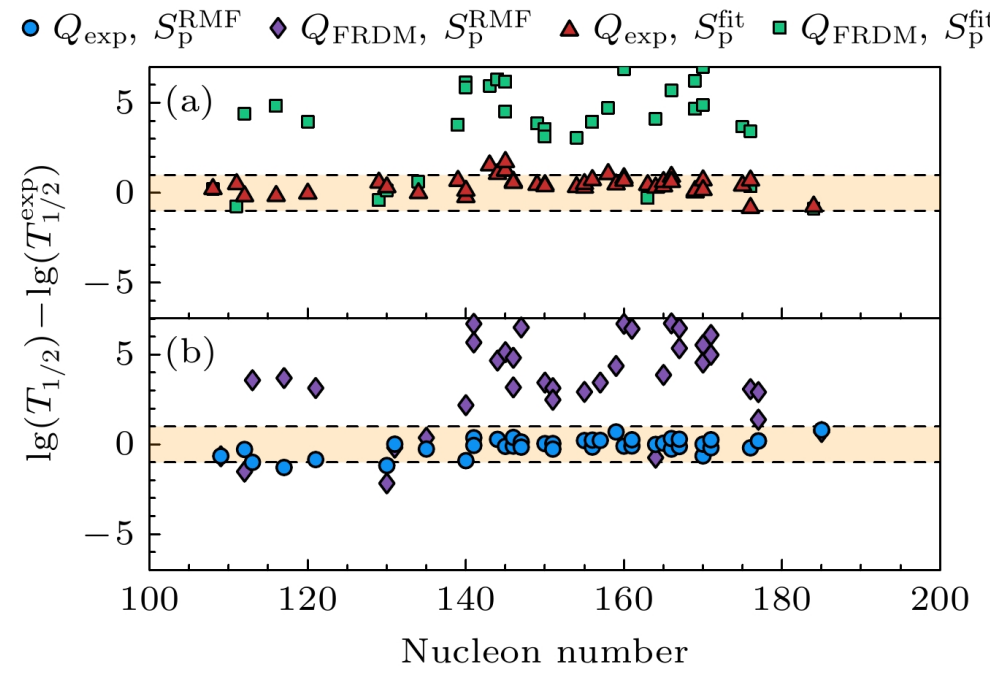

**Fig. 3 Errors of the theoretical model in experimental and FRDM decay energies: (a) Using the spectroscopic factor obtained by fitting; (b) using the spectroscopic factor derived from RMF theory. The shaded region indicates that the absolute difference between our calculated value and the experimental value, under logarithmic conditions, is less than 1.**

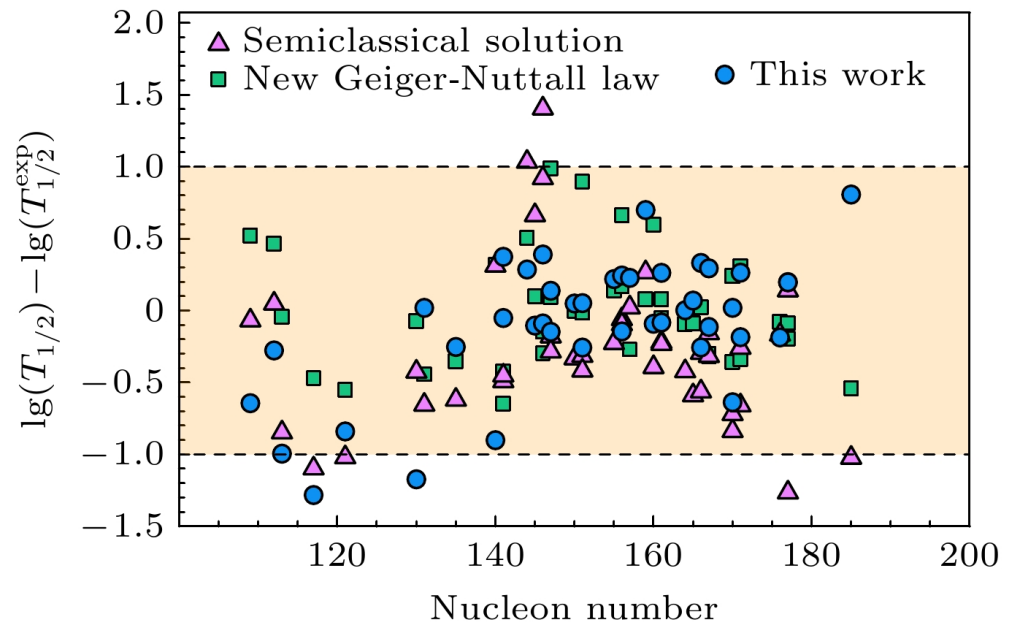

**Fig. 4 The difference between this work and other theoretical methods (including an analytic semiclassical solution[11] and the new Geiger-Nuttall law[44]). The horizontal axis denotes the nucleon number of the parent nucleus, while the vertical axis illustrates the difference between the logarithms of the calculated and experimental half-lives.**

## 3.2 Theoretical Predictions for Lighter Proton Radioactive Nuclei

In recent years, exploration of nuclei beyond the proton drip line has led to the discovery of an increasing number of lighter, proton-unbound nuclei, extending into the region where $Z < 50$. However, the half-lives of numerous potential proton emitters in this region remain experimentally undetermined[47]. Consequently, theoretical calculations of half-lives for nuclei in the $Z < 50$ region hold significant value for guiding future experimental investigations into the properties of these nuclei with extreme neutron-to-proton ratios. Taking $^{89}$Rh-a nucleus whose proton radioactivity has been experimentally confirmed-as an example, the emitted proton originates from an orbital with $\ell = 4, j^{\pi} = 9/2^{+}$. Experimental data indicate a decay energy below 1 MeV[47]; thus, the present study adopts the FRDM theoretical value of 0.964 MeV as the decay energy. The RMF theory yields a spectroscopic factor of 0.56 for $^{89}$Rh[48], whereas a value of 0.158 is obtained from deformation-parameter fitting[13]. Using these two spectroscopic factors, our theoretical model predicts half-lives of 274 ns and 972 ns, respectively-both on the sub-microsecond scale. This contrasts markedly with the heavier proton emitters discussed in Section 3.1, reflecting distinct properties of proton radioactivity across different nuclear shells.

Subsequently, to conduct a more comprehensive investigation of lighter proton emitters below the $Z = 50$ magic number, we performed calculations for additional odd-$Z$ proton emitters within the *fpg* shell, spanning $Z = 30$ to $Z = 50$. These calculations combine single-proton separation energies from the NNDC database[3] with prior theoretical predictions of the proton drip line in this region[48]. As established in Section 3.1, to achieve maximal precision in half-life calculations, we use experimental single-proton separation energies from NNDC as the proton decay energy $Q$[3]. For nuclei lacking experimental data, the database provides decay energies derived from the AME2020 mass table using TMS (trends from the mass surface) mass evaluations. In our data tables, such evaluated decay energies are marked with a superscript #[40,41]. Although spectroscopic factors from RMF calculations and those obtained via fitting exhibit comparable accuracy in half-life predictions, the proton emitters considered here reside in different nuclear shells. Therefore, we do not employ the fitting formula previously derived from data in the $Z = 50$ to $Z = 84$ region. Instead, we adopt spectroscopic factors systematically computed using the relativistic Hartree-Bogoliubov (RHB) model[48]. All other nuclear parameters follow the same sources as in Section 3.1. Proton orbitals in the spherical limit are determined based on Nilsson levels within this shell and available spin-parity assignments[3,38]. **Table 3** presents the data and theoretical predictions for possible proton emitters in the *fpg* shell. The first two columns list the parent nucleus and the orbital of the emitted proton. The third column provides the decay energy $Q$, calculated from experimental single-proton separation energies in the NNDC database and the AME2020 mass table[3,40,41] (values marked with # denote AME2020 evaluations). Columns 4 and 5 list the quadrupole and hexadecapole deformations ($\ell = 1$) of the daughter nucleus ground state[39]. Column 6 gives the spectroscopic factor from RHB calculations. The final column shows the logarithm of the proton radioactivity half-life (in seconds) predicted in this work. To illustrate the distribution pattern of these predicted half-lives, Figure 5 plots the logarithm of the predicted proton radioactivity half-life against the proton number and decay energy of the parent nucleus. In the *fpg* shell, quasi-bound protons may be emitted from states with $\ell = 1$, 3, or 4, represented by distinct symbols in the figure. For proton emitters with identical angular momentum, a clear dependence between half-life and decay energy emerges. Owing to analogous physical mechanisms, the Geiger-Nuttall law-well established for α decay-also applies to proton radioactivity. Accordingly, this work follows the half-life-decay energy relationship prescribed by this law.

**Table 3 The theoretical prediction on the half-lives of fpg-shell possible proton emitters. The decay energy $Q$ is taken from the proton separation energy in NNDC and AME 2020[3,40,41] (The superscript # denotes values that are not obtained from purely experimental data in AME 2020), and the deformation parameters $\beta_{2,4}$ and spectroscopic factor $S_{\mathrm{p}}^{\mathrm{RHB}}$ are obtained by the results of FRDM[39] and RHB[48], respectively. The predicted half-lives (unit: s) using our model are listed in the last column.**

| Parent Nuclei | Orbital | $Q$/MeV | $\beta_2$ | $\beta_4$ | $S_{\mathrm{p}}^{\mathrm{RHB}}$ | $\log_{10} T_{1/2}^{\mathrm{pred}}$ |
|---|---|---|---|---|---|---|
| $^{60}$Ga | $2p_{3/2}$ | 0.340# | 0.106 | 0.041 | 0.53 | −5.140 |
| $^{63}$As | $2p_{3/2}$ | 1.350# | 0.184 | 0.014 | 0.61 | −15.627 |
| $^{68}$Br | $1f_{5/2}$ | 0.500# | 0.220 | −0.082 | 0.80 | −5.071 |
| $^{69}$Br | $1f_{5/2}$ | 0.640 | 0.233 | −0.106 | 0.78 | −7.489 |
| $^{72}$Rb | $1f_{5/2}$ | 0.710# | −0.357 | 0.022 | 0.83 | −7.713 |
| $^{73}$Rb | $1f_{5/2}$ | 0.640 | −0.366 | 0.025 | 0.83 | −6.751 |
| $^{75}$Y | $1g_{9/2}$ | 1.720# | 0.401 | 0.001 | 0.92 | −12.882 |
| $^{76}$Y | $1g_{9/2}$ | 1.080# | 0.402 | −0.012 | 0.84 | −9.505 |
| $^{81}$Nb | $1g_{9/2}$ | 1.110# | 0.430 | −0.047 | 0.12 | −8.391 |
| $^{84}$Tc | $1f_{5/2}$ | 1.350# | 0.492 | −0.021 | 0.90 | −11.544 |
| $^{88}$Rh | $1g_{9/2}$ | 1.580# | −0.243 | −0.101 | 0.73 | −10.976 |
| $^{92}$Ag | $1g_{9/2}$ | 1.350# | −0.011 | 0.000 | 0.56 | −9.074 |
| $^{93}$Ag | $1g_{9/2}$ | 1.090# | 0.000 | 0.000 | 0.39 | −6.967 |
| $^{96}$In | $1g_{9/2}$ | 1.680# | 0.053 | 0.001 | 0.49 | −10.374 |
| $^{97}$In | $1g_{9/2}$ | 0.890# | −0.021 | 0.000 | 0.19 | −3.965 |

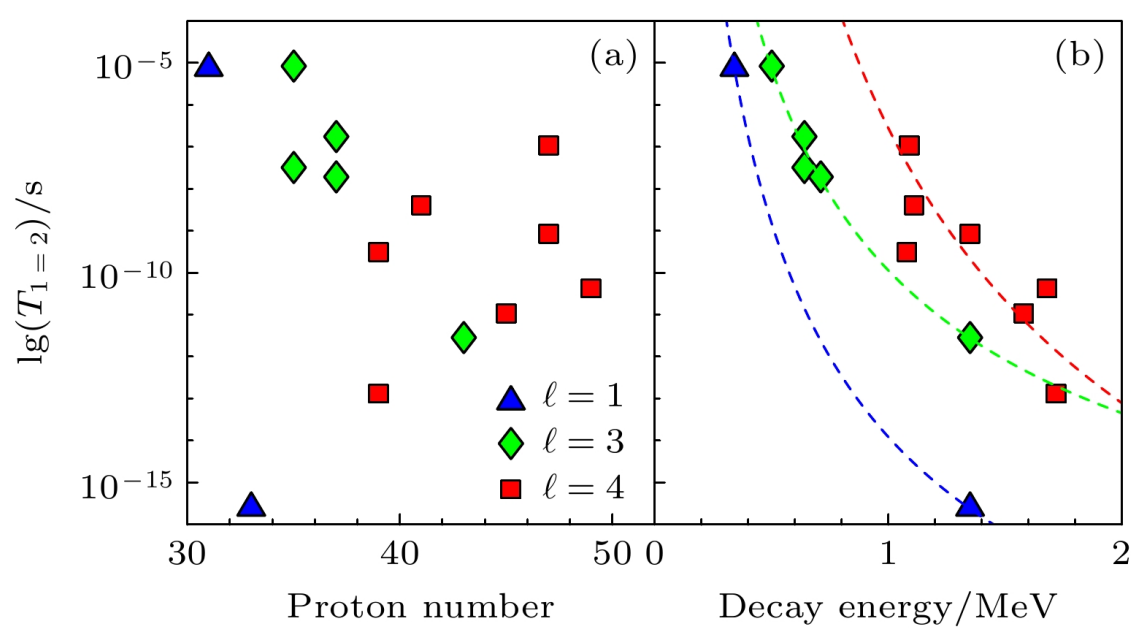

**Fig. 5 The predicted half-lives for serial proton emitters in the range from $Z = 30$ to $Z = 50$ in this work, and the dependence on (a) the proton number of parent nuclei and (b) the decay energy, which is consistent with the Geiger-Nuttall law as indicated by the dashed lines.**

$$\log_{10} T_{1/2} = aQ^{-\frac{1}{2}} + b, \tag{10}$$

We performed separate fits for the predicted proton-emitting nuclei with different $\ell$ values. The resulting fits are shown as the dashed lines in Fig. 5(b). For protons with the same angular momentum, the probability of resonant-state proton tunneling through the potential barrier increases as the decay energy rises, leading to shorter half-lives. This fitting result indicates that proton radioactivity shares a similar physical mechanism with α decay. Moreover, it is noteworthy that, compared with the experimentally observed proton emitters listed in Table 1, the lighter proton-emitting nuclei predicted in this work within the fpg shell exhibit relatively shorter half-lives, ranging from $10^{-4}$ to $10^{-16}$s. This timescale poses significant challenges for conventional radioactivity measurement techniques. However, recent experimental investigations of additional light proton-unbound nuclei have expanded the measurable half-life range of traditional methods and enhanced experimental sensitivity. Techniques such as in-flight decay measurements[49,50] are now capable of probing half-lives on the nanosecond (ns) to picosecond (ps) scale, offering promising avenues for further studies of the structure and radioactive behavior of extremely proton-rich nuclei with even shorter lifetimes.

# 4 Conclusion

In summary, this work first reviewed the quantum tunneling model for proton radioactivity under deformed conditions. Subsequently, we employed the Woods-Saxon nuclear potential combined with spin-orbit interaction and the Coulomb potential to calculate half-lives for a large number of proton-emitting nuclei already observed experimentally, comparing our results with experimental values. To analyze the influence of different nuclear inputs on the calculations, we separately used experimental decay energies and decay energies derived from the Finite-Range

Droplet Model (FRDM), as well as spectroscopic factors from Relativistic Mean-Field (RMF) theory and deformation-dependent fitted spectroscopic factors. The results demonstrate that proton radioactivity is highly sensitive to decay energy; calculations using experimental decay energies show good agreement with experimental data. Furthermore, based on existing experimental information, we theoretically predicted the half-life of the lighter proton emitter $^{89}$Rh, which lies in the shell below the one under study and has already been experimentally identified. Using the two types of spectroscopic factors, the predicted half-lives for $^{89}$Rh are 274 ns and 972 ns, respectively. Additionally, we made theoretical predictions for the proton radioactivity half-lives of other proton-unbound nuclei with proton numbers between 30 and 50, finding that they generally possess shorter half-lives compared to the heavier proton emitters currently known. Finally, we systematically compiled a comprehensive dataset of theoretically calculated and predicted proton radioactivity half-lives from this work, spanning nuclei from $^{60}$Ga to $^{185}$Bi. We hope this study provides a valuable theoretical reference for future experiments on nuclei near the proton drip line.

## Data Availability Statement

The dataset supporting the findings of this study is accessible at the Science Data Bank: https://www.doi.org/10.57760/sciencedb.27551.